# Submerged turbulence detection with optical satellites

Carl H. Gibson[a], R. Norris Keeler[b], Valery G. Bondur[c], Pak T. Leung[d], H. Prandke[e], D. Vithanage[f]

**ABSTRACT**

During fall periods in 2002, 2003 and 2004 three major oceanographic expeditions were carried out in Mamala Bay, Hawaii. These were part of the RASP Remote Anthropogenic Sensing Program. Ikonos and Quickbird optical satellite images of sea surface glint revealed $\approx 100$ m spectral anomalies in $km^2$ averaging patches in regions leading from the Honolulu Sand Island Municipal Outfall diffuser to distances up to 20 km. To determine the mechanisms behind this phenomenon, the RASP expeditions monitored the waters adjacent to the outfall with an array of hydrographic, optical and turbulence microstructure sensors in anomaly and ambient background regions. Drogue tracks and mean turbulence parameters for $2 \times 10^4$ microstructure patches were analyzed to understand complex turbulence, fossil turbulence and zombie turbulence near-vertical internal wave transport processes. The dominant mechanism appears to be generic to stratified natural fluids including planet and star atmospheres and is termed beamed zombie turbulence maser action (BZTMA). Most of the bottom turbulent kinetic energy is converted to $\approx 100$ m fossil turbulence waves. These activate secondary (zombie) turbulence in outfall fossil turbulence patches that transmit heat, mass, chemical species, momentum and information vertically to the sea surface for detection in an efficient maser action. The transport is beamed in intermittent mixing chimneys.

**Keywords:** Turbulence-stratified ocean, satellite remote sensing, fossil turbulence waves, zombie turbulence waves

## 1. INTRODUCTION

Astronauts have remarked that oceanic bottom features such as seamounts and bottom topography can often be observed from space. Synthetic aperture radar images show surprising details of deeply submerged turbulence and internal wave patterns. How is this possible? Because the ocean is almost everywhere stably stratified, turbulence∗ is strongly inhibited by buoyancy forces and will fossilize∗ when it cascades from small Kolmogorov inertial-viscous scales to larger Ozmidov fossil turbulence inertial-buoyancy scales[1]. Most of the kinetic energy of turbulence events in a stratified medium is converted to fossil vorticity turbulence kinetic energy and radiated near-vertically as fossil turbulence internal waves[2]. The hydrodynamic state of stratified oceanic microstructure patches is determined using hydrodynamic phase diagrams[3]. Figure 1 summarizes the method and results and illustrates the ships, satellites and sensors used in the three RASP expeditions, as reported elsewhere[4,5]. See also Keeler, R.N. V. G. Bondur, and D. Vithanage, Sea Technology 45(4), 53(2004) and Wolk, Fabian, Carl H. Gibson and Hartmut Prandtke, Sea Technology 45(8), 47(2004).

---

[a] Departments of Mechanical and Aerospace Engineering and Scripps Institution of Oceanography, University of California at San Diego, La Jolla, CA, 92093-0411, USA, http://sdcc3.ucsd.edu/~ir118, cgibson@ucsd.edu
[b] Directed Technologies, Inc., Arlington 22201, VA, USA, norris_keeler@directedtechnologies.com
[c] Aerocosmos Scientific Ctr. of Aerospace Monitoring, Russia, vgbondur@online.ru
[d] Dept. of Oceanogr., Phys. Sect., Texas A&M Univ., College Station, TX 77843, USA, ptleung@tamu.edu
[e] ISW Wassermesstechnik, Petersdorf, Germany, prandke@t-online.de
[f] Oceanit Labs., Inc., Honolulu, HI, USA, dvithanage@oceanit.com

∗ Turbulence is defined as an eddy-like state of fluid motion where the inertial-vortex forces $\vec{v} \times \vec{\omega}$ of the eddies are larger than any other forces that tend to damp the eddies out, where $\vec{v}$ is the velocity and $\vec{\omega}$ is the vorticity. Thus, turbulence always cascades from small scales to large.
∗ Fossil turbulence is defined as a perturbation in any hydrophysical field produced by turbulence that persists after the fluid ceases to be turbulent at the scale of the perturbation.

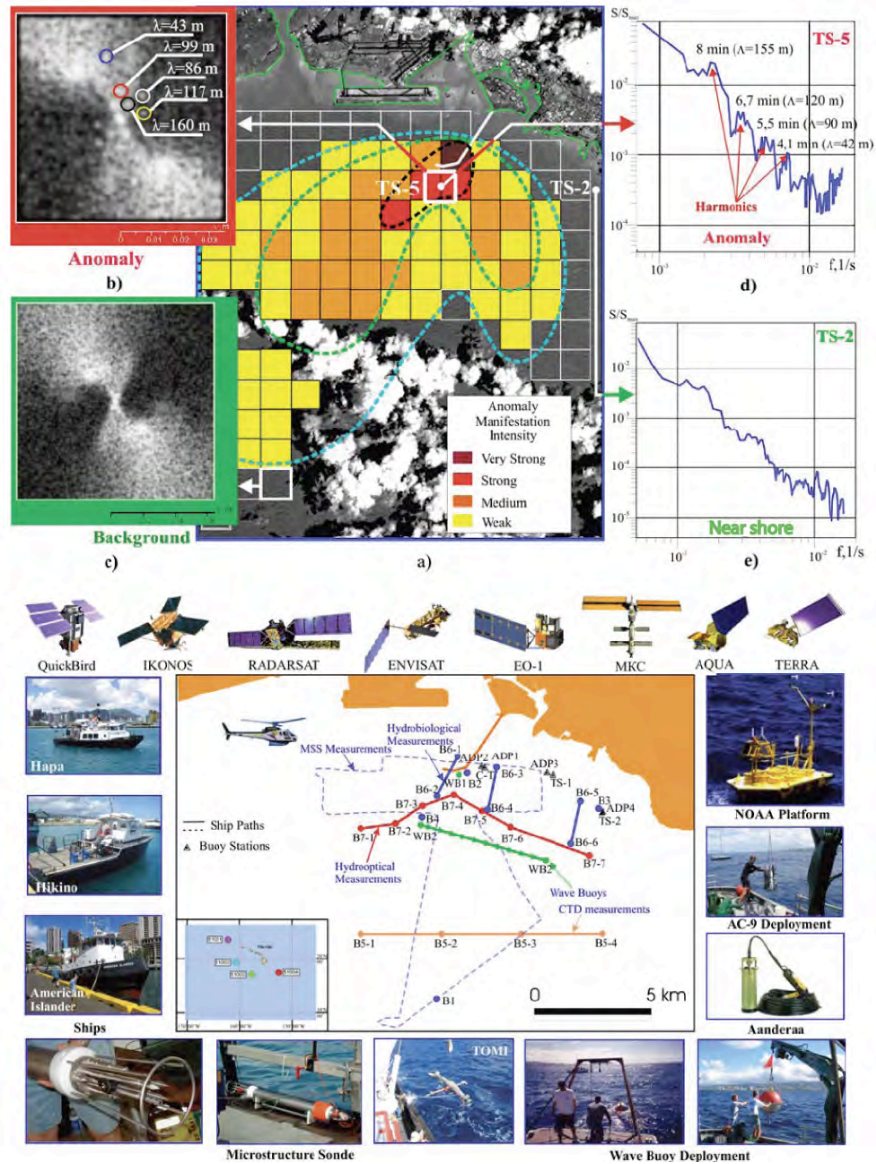

Figure 1. Sea surface brightness anomaly map for Sept. 4, 2004, QuickBird image and internal wave spectra from thermistor strings TS-5 and TS-2. Wavelengths λ of the anomalies match wavelengths Λ of the internal waves at TS-5's position just south of the diffuser and at TS-2 nearer shore (top)[5]. The complex array of ships, satellites, microstructure detectors, platforms, and section paths used in RASP is shown (bottom).

Figure 2 shows the range of anomaly regions for the RASP experiments, as well as a synthetic aperture radar map for anomalies in the region above Hudson Canyon in the New York Bight where we infer that large fossil turbulence patches from the Canyon amplify Beamed Zombie Turbulence Maser Action (BZTMA) surface manifestations of submerged turbulence produced by tidal internal solitary waves (ISWs).

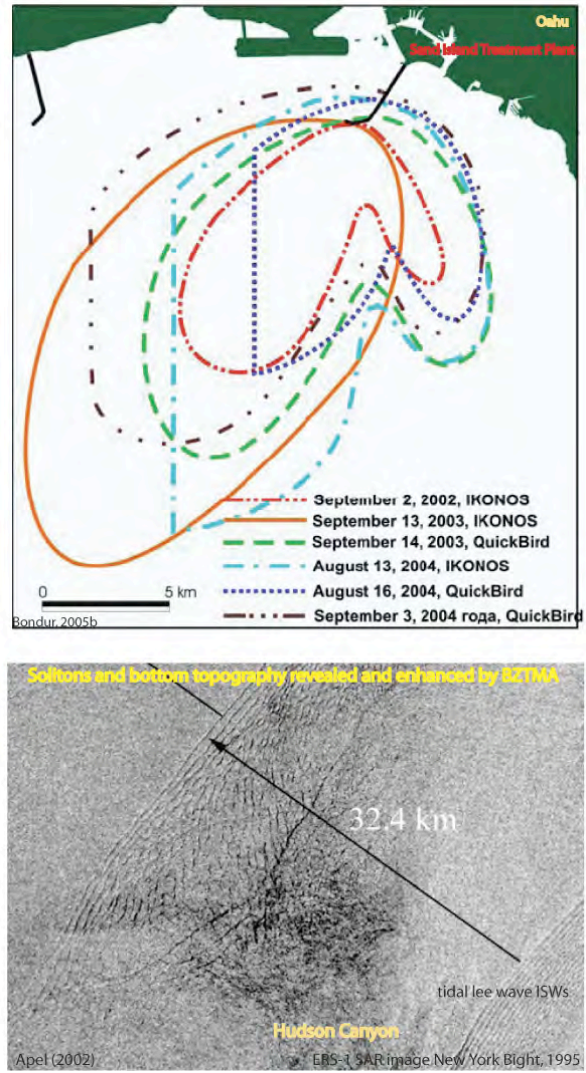

Figure 2. (top) Spectral quasimonochromatic anomaly regions (λ wavelengths 30-250 m) detected from optical satellite images during the remote anthropogenic sensing program RASP 2002, 2003, 2004[5,6]. The anomaly regions reflect enhancement of the BZTMA mechanism by outfall fossil turbulence patches. (bottom) Radar image showing that surface manifestations of tidal internal solitary waves and several fronts enhanced over Hudson Canyon by turbulence patches from fossil turbulence waves radiated near vertically from strong bottom turbulence generated in the canyon.

## 2. THEORY

Figure 3 shows schematically the turbulence, fossil turbulence and fossil turbulence wave mechanisms inferred from the RASP experiments for the remote detection of Sand Island Outfall submerged turbulence. Because the vertical velocity of the fossil turbulence waves (FTWs) matches the horizontal velocity of the turbulence source, the angle of propagation is a near-vertical $45°$. Because a substantial fraction of the turbulent kinetic energy is radiated near-vertically in narrow-wavenumber-band FTWs with $\lambda \approx L_{R_0}$ this is an efficient hydrodynamic maser-action. The presence of ambient fossils of strong turbulence enhances the vertical information transfer to produce the anomaly regions of Fig. 1 and Fig. 2.

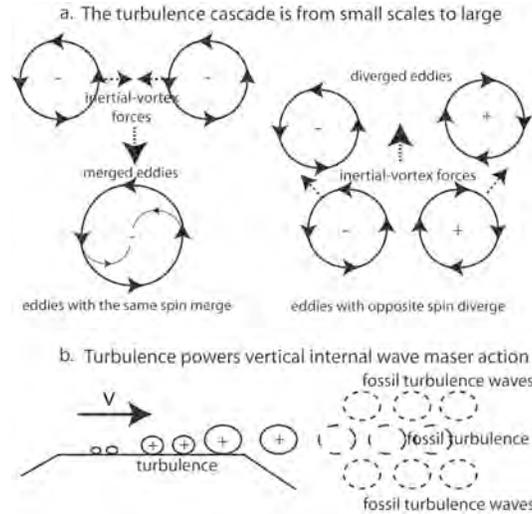

Figure 3. Physical mechanisms of turbulence and stratified turbulence. a. Vortex mechanisms of the turbulence cascade from small scales to large. Adjacent eddies with the same vorticity produce inertial vortex forces $\vec{v} \times \vec{\omega}$ (dashed arrows) that cause merging. Nearby eddies with opposite spin diverge and expand the turbulent region driven by $\vec{v} \times \vec{\omega}$ forces. b. Turbulence, fossil turbulence, and fossil-turbulence-waves in a stratified fluid produce internal-wave maser-action where turbulent kinetic energy fossilized by buoyancy forces is radiated near vertically as fossil turbulence waves (FTWs).

Figure 4 summarizes the remote detection mechanisms for the Sand Island Outfall submerged turbulence.

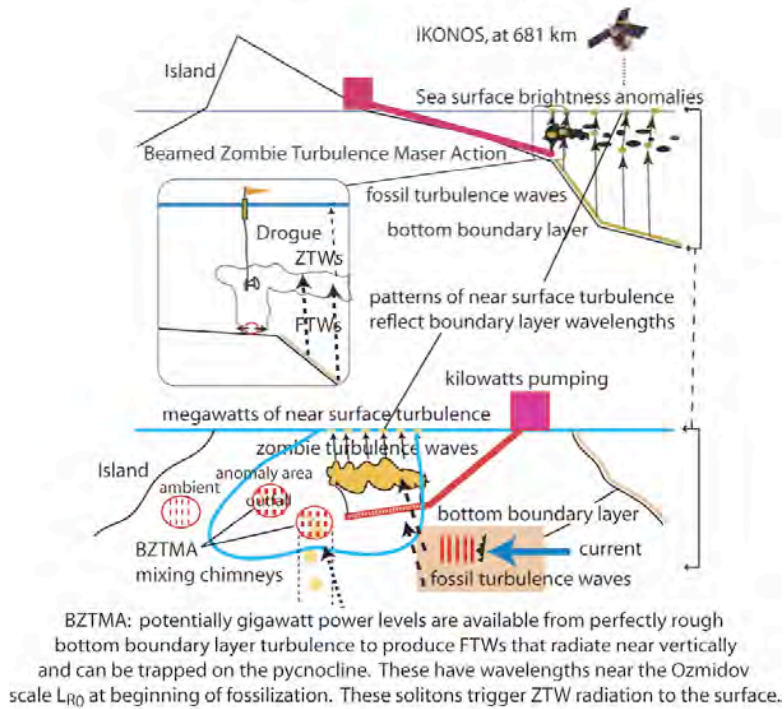

Figure 4. Beamed zombie turbulence maser action (BZTMA) model[2,4,6] for the RASP remote detection of submerged outfall turbulence and fossil turbulence. Buoyancy traps the outfall wastefield about 50 meters below the surface. Spectral anomalies with precise wavelengths in the range 43-160 meters are detected near the outfall diffuser, as shown in Fig. 1, but not in ambient background regions. The anomalies are attributed to bottom generated FTWs interacting with outfall fossil turbulence patches to produce ZTWs that radiate the information to the sea surface where the patterns also fossilize and are detected.

## 3. RESULTS

Figure 5 shows sample results of the study for Sept. 13, 2003, when the largest anomaly region of Fig. 2 was detected. Vertical and horizontal profiling of microstructure revealed evidence of zombie turbulence mixing chimneys, as well as a mixing front reflecting powerful BZTMA mixing of Mamala Bay by outfall fossils advected far off shore by previous rain induced horizontal transport[5,6]. GPS equipped parachute drogues tracked currents convecting outfall fossil turbulence patches off shore in the patterns of the anomaly regions.

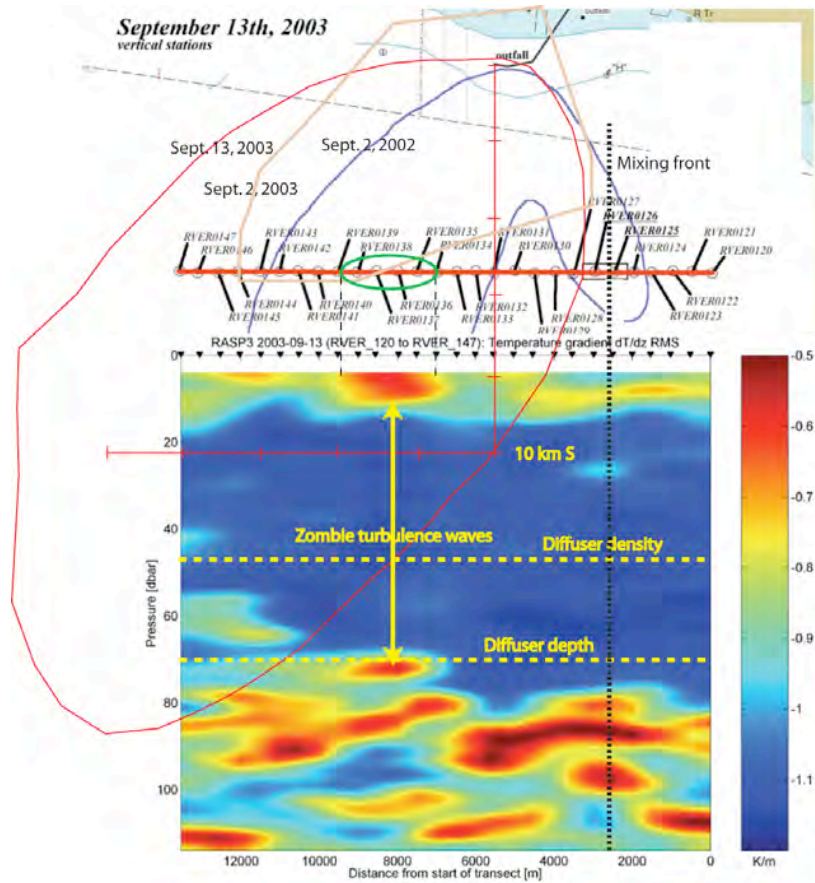

Figure 5. Contour plot of the Sept. 13, 2003 rms temperature dissipation rate (as log (dT/dz)rms) for vertical profiles 5.5 km south of the diffuser. Active near surface regions of this indicator of fossil temperature turbulence are interpreted as evidence of breaking zombie turbulence waves (double arrows) above the wastewater trapping depth. The Sept. 13, 2003 area of surface brightness anomalies is about 200 km$^2$, ~ 4 times larger than that for Sept. 2, 2002.

Figure 6 shows a collection of several thousand hydrodynamic phase diagrams (HPDs) from RASP 2002 and RASP 2003. Viscous dissipation rates $\varepsilon$ are measured with shear probes on the dropsonde for a microstructure patch and normalized with $\varepsilon_0$ and $\varepsilon_F$ to determine the normalized Froude number and normalized Reynolds number, where $\varepsilon_0$ is the dissipation rate of the patch at the beginning of fossilization (estimated from the overturn scale of the patch and the ambient stratification $N$) and $\varepsilon_F = 30\nu N^2$ is the dissipation rate at complete fossilization, where $\nu$ is the kinematic viscosity (see references 1-5). Most patches were in the active-fossil quadrant, where only the smaller scale eddies are overturning and fully turbulent. Even though hundreds of profiles were taken near the outfall, only three outfall patches were found in their fully turbulent state, showing the difficulty of adequately sampling the highly intermittent stratified turbulence physical process and the importance of using HPDs to properly interpret oceanic microstructure data.

Zombie turbulence outfall patches are identified by their position on decay-activation lines of slope +1/3 slightly below the locus for outfall patches, and always in the active-fossil turbulence quadrant.

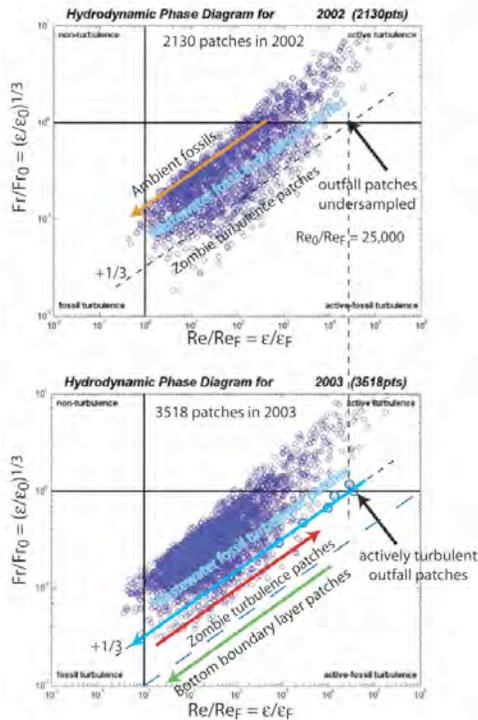

Figure 6. Hydrodynamic phase diagram points for all patches detected in RASP 2002 and RASP 2003 experiments[6]. The patterns suggest the wastewater outfall produces dominant patches with $Re_0/Re_F$ values of 25,000 significantly larger than ambient (orange arrow, $Re_0/Re_F \sim 10^3$). Patches to the right of the +1/3 locus for this $Re_0/Re_F$ are interpreted as zombie turbulence (red arrow, $Re_0/Re_F \sim 10^5$). These beam information to the sea surface about the wavelength and strength of the internal waves from which they extract energy according to the BZTMA model. Deep measurements in RASP 2004 revealed fossils of BBL turbulent events with $Re_0/Re_F$ values of $10^6$.

Figure 7 shows an astrophysical application of BZTMA mixing theory[7]. Radio telescope measurements of electron density spectra in local regions of the Galaxy exhibit the Kolmogorovian universal similarity forms of turbulent mixing. Without BZTMA, a new fluid-mechanically-based cosmology (termed Hydro-Gravitational-Dynamics, HGD) and uniform supernova this "great power law on the sky" is a mystery. Uniform pulsar scintillation spectra in all directions indicate a transition from the -11/3 slope expected for Obukov-Corrsin turbulent mixing to the -15/3 slope expected from the Gibson 1968ab mixing theory for such strongly diffusive scalars with diffusivity $D \approx 30\nu$. For turbulence or fossils of turbulent mixing to exist in the near vacuum of space requires the existence of dense planetary atmospheres, but these require evaporated gas planets. Frozen primordial gas planets (JPPs) in proto-globular-star-cluster (PGC) clumps have been proposed[7] (Gibson-Schild 1996) as the dark-matter of galaxies formed soon after the big bang, where $m_{galaxy} = m_{Planets} + m_{Stars}$, $m_{Planets} \approx 30 m_{Stars}$. All stars form and grow from dark-matter planets. When stars explode to form spinning neutron stars they evaporate some of the $3 \times 10^7$ interstellar planets per star to form strong turbulence in stably stratified atmospheres detected as electron-density-produced scintillations of the pulsar signals. Mixing of the stratified atmospheres over the observed range of scales can only be understood using BZTMA mixing theory.

The largest scales of the spectra correspond to the $3 \times 10^{17}$ m size of a PGC clump, each containing a trillion rogue planets, normally identified as a giant molecular cloud. More than a million planets merge to form a star, leaving a $3 \times 10^{15}$ m Oort cavity. Thousands of evaporating $10^{13}$ m JPPs surround white dwarf stars in "planetary nebulae" such as the nearby Helix, detected by the Hubble Space Telescope. Evaporated JPP atmospheres and wakes are also seen in supernovae remnants such as the Crab, with planet wakes tracking the motion of the resulting pulsar. Powerful spin powered equatorial radiation from white dwarfs and neutron stars can only be explained as the result of stratified turbulent mixing of stratified plasma and magnetic fields by beamed zombie turbulence maser action (BZTMA).

The main difference between formation of white dwarfs and neutron stars is the mass mixing rates of dark matter JPP planets. These stars collapse, spin, and evaporate ambient planets by resulting axial plasma jets and equatorial turbulent plasma winds to form planetary nebulae. If the planet accretion rate is very slow, compact helium or carbon stars may cease growing and cool from lack of fuel. If the accretion rate is moderate the compressing carbon core may reach Chandrasekhar instability at 1.44 solar star mass and explode as a supernova type Ia event. If the accretion rate is higher then BZTMA mixing entrains and burns the carbon core of the star so elements with larger atomic number can form up to iron, which then may collapse to a spinning neutron star (pulsar) at 1.4 times solar mass in a supernova type IIab event. Higher JPP mixing rates may produce superstars and black holes. Because neutron stars have densities of $\approx 10^{18}$ kg m$^{-3}$, equatorial spin speeds approach the speed of light creating powerful stratified turbulent mixing and BZTMA equatorial radiation by the mixed magnetic field. This and the supernova II radiation account for the evaporation of ambient JPPs and the 100% intersection of numerous pulsar lines of sight in Fig. 7 with turbulent or fossil turbulence planetary atmospheres. White dwarfs have densities of only $\approx 10^{10}$ kg m$^{-3}$ so lines-of-sight to supernovae Ia are only intermittently dimmed by dark matter planet atmospheres. This intermittent dimming has been widely misinterpreted as an accelerating (rather than decelerating) expansion rate of the universe driven by dark energy[7].

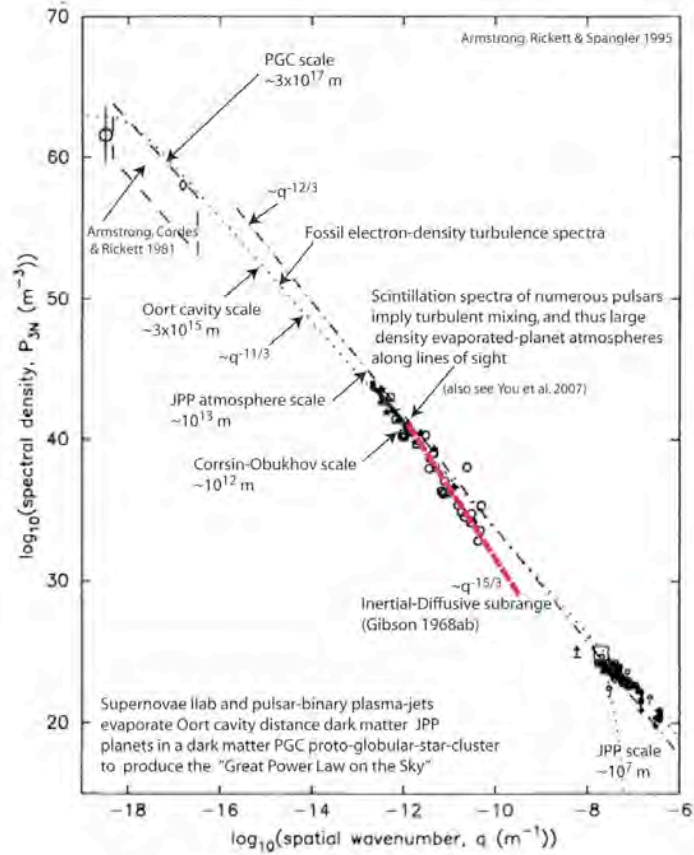

Figure 7. Application of BZTMA mixing theory to understand pulsar electron density fluctuation spectra and star formation from planets[7]. Jovian PFP (primordial –fog-particle) Planets (JPPs) comprise the baryonic dark matter of all galaxies and develop turbulent atmospheres when evaporated by radiation from rapidly spinning white dwarf and neutron stars.

## 4.  CONCLUSIONS

Microstructure and internal-wave measurements from vertical and horizontal profilers near a Honolulu municipal wastewater outfall are compared to soliton-induced sea-surface brightness anomalies from optical and synthetic-aperture-radar space satellite images.  Spectral anomalies with wavelengths 30-1000 m were detected September 2, 2002. Anomaly areas covered 70 km$^2$ in 10 km and 5 km SW and SE lobes extending from the diffuser. Studies in 2003 and 2004 increase the range of detection to 20 km from the outfall in areas covering 200 km$^2$.  The remote detection mechanism indicated by these remarkable observations is a complex interaction between advected 10 m outfall fossil turbulence patches and internal-soliton-waves (ISWs). ISWs supply new turbulent-kinetic-energy to outfall patches near the pycnocline depth.  Energy, mixing, and information is radiated near-vertically by both the primary and secondary fossil turbulence patches in ISW patterns of surface smoothing, as detected from space.  Nonlinear vertical-amplification and vertical-beaming internal wave processes are similar to those of astrophysical masers but more efficient.  Off shore advection of the outfall fossil turbulence patches required to produce the anomaly lobes varies widely and unpredictably, and depends on fresh water run off from the island.  The identified beamed zombie turbulence maser action (BZTMA) mixing chimney process appears to be genetic to vertical mixing and transport of heat, mass, momentum and chemical species in the ocean and atmosphere, with applications in astronomy, astrophysics and cosmology.

## REFERENCES

[1] Gibson, C.H. 1986.  Internal waves, fossil turbulence, and composite ocean microstructure spectra, *J. Fluid Mech.*, 168, 89-117.

[2] Gibson, C.H., Bondur, V.G., Keeler, R.N. & Leung, P.T. 2006.  Remote sensing of submerged oceanic turbulence and fossil turbulence, *Int. J. Dyn. Fluids,*, Vol. 2, No. 2, pp. 111-135.

[3] Leung, P. T. and Gibson, C. H. 2004.  Turbulence and fossil turbulence in oceans and lakes, Chinese Journal of Oceanology and Limnology, 22(1), 1-23. http:// xxx.lanl.gov, astro-ph/ 0310101.

[4] Keeler, R. N., Bondur, V. G., and Gibson, C. H. 2005.  Optical satellite imagery detection of internal wave effects from a submerged turbulent outfall in the stratified ocean, Geophys. Res. Lett., 32, L12610, doi:10.1029/2005GL022390.

[5] Bondur V.G., Complex satellite monitoring of coastal water areas 2005.  31st International Symposium on Remote Sensing of Environment. ISRSE June 20-24, St. Petersburg, Russian Federation.

[6] Gibson, C.H., Bondur, V.G., Keeler, R.N. & Leung, P.T. 2006.  Energetics of the beamed zombie turbulence maser action mechanism for remote detection of submerged oceanic turbulence, J. Appl. Fluid Mech., Vol. 1, No. 1, pp. 11-42.

[7] Gibson, C. H. & Schild, R. E. 2007.  Interpretation of the Helix planetary nebula using hydro-gravitational-dynamics: planets and dark energy, astro-ph/0701474.